\documentclass{article}
\usepackage{graphicx}
\usepackage{tabularx}

\begin{document}

\author{Salman Khan$^{\dag }$\thanks{%
sksafi@phys.qau.edu.pk} M. K. Khan$^{\ddag }$ \\
$^{\dag }$ Department of Physics, COMSATS Institute of Information \\
Technology, Islamabad, Pakistan\\
$^{\ddag }$ Department of Physics, Quaid-i-Azam University,\\
Islamabad, Pakistan}
\title{Entanglement of Open Quantum Systems in Noninertial Frames}
\maketitle

\begin{abstract}
We study the effects of decoherence on the entanglement generated by Unruh
effect in accelerated frames by using various combinations of an amplitude
damping channel, a phase damping channel and a depolarizing channel in the
form of multilocal and collective environments. Using concurrence as
entanglement quantifier, we show that the occurrence of entanglement sudden
death (ESD) depends on different combinations of the channels. The ESD can
be avoided under a particular configuration of the channels. We show that
the channels can be used to distinguish between a moving and a stationary
frame.\newline

\end{abstract}

\section{Introduction}

The last two decades has witnessed the rapid development in the field of
quantum information and strengthened the notion that entanglement is not
only one of the fundamental concepts of quantum theory but also a key
concept for transmitting and processing quantum information \cite{springer}.
Entanglement between spatially separated parties is used as a potential
source for quantum teleportation of unknown states \cite{Bennett}, quantum
key distribution \cite{Ekert}, quantum cryptography \cite{Bennett2} and
quantum computation \cite{Grover, Vincenzo}. However, the behavior of
entanglement between various systems is still not fully known and efforts
are on the line to understand its dynamics deeper under various setups. The
study of entanglement in a bipartite system has recently been extended to
the relativistic setup and its behavior for various fields has been examined
\cite{Alsing,Ling,Gingrich,Pan, Schuller, Terashima}. However, these
investigations on entanglement are limited to closed quantum systems.
Practically, quantum systems are influenced by its environment that may
results in non-unitary dynamics of the system. For most accurate and
practical results in quantum processing, the effect of environment on the
entanglement between spatially separated systems needs to be necessarily
investigated. The environmental effect on a quantum system gives rise to the
phenomenon of decoherence that causes an irreversible transfer of
information from the system to the environment \cite{Zurik, Breuer, Zurik2}.

Alsing \textit{et al} \cite{Alsing} have shown that the entanglement between
two modes of a free Dirac field is degraded by the Unruh effect and
asymptotically reaches a nonvanishing minimum value in the infinite
acceleration. In Ref. \cite{Wang}, the dynamics of entanglement for Dirac
field in the presence of Unruh effect has been studied under amplitude
damping channel. The effect of decoherence on the entanglement of Dirac
field using a phase damping channel, a depolarizing channel and a phase flip
channel has been investigated in Ref. \cite{Salman}.

In this paper we investigate the effect of decoherence on the entanglement
of Dirac field in a noninertial system by considering various combinations
of an amplitude damping, a phase damping and a depolarizing channel in the
form of multilocal, collective and global noises. For example, we consider
that if one qubit is locally coupled to an amplitude damping channel and the
other is locally coupled to a phase damping channel then both the qubits are
collectively coupled to a depolarizing channel. Using concurrence as the
entanglement quantifier, we show that the rate of decrease of concurrence
depends on the coupling of the kind of a channel to a particular qubit. For
example, under the influence of multilocal coupling when Alice's qubit is
influenced by phase damping channel and Rob's qubit is influenced by
amplitude damping channel, the entanglement sudden death (ESD) \cite{Yu}
occurs earlier for large acceleration of the frames under some particular
conditions. On the other hand, Under some other setups, the ESD can be
avoided or delayed or even faster.

We rigorously consider three different cases. In one case we allow Rob's
qubit to interact locally with an amplitude damping channel and Alice's
qubit interact locally with a phase damping channel. In the second case, the
coupling of these two channels are reversed in the sense that Rob's qubit
interacts with a phase damping channel and Alice's qubit interacts with an
amplitude damping channel. However, in both these cases the two qubits are
coupled collectively with a depolarizing channel. In the third case, the two
channels in the multilocal coupling of the qubits are depolarizing and phase
damping channels while collectively the two qubits are coupled to an
amplitude damping channel.

We consider that Alice and Rob, the two parties, share the following
maximally entangled state at a point in flat Minkowski spacetime%
\begin{equation}
|\psi \rangle _{M}=\frac{1}{\sqrt{2}}\left( |0_{\omega _{A}}\rangle
_{M}|0_{\omega _{R}}\rangle _{M}+|1_{\omega _{A}}\rangle _{M}|1_{\omega
_{R}}\rangle _{M}\right) .  \label{1}
\end{equation}%
In Eq. (\ref{1}) $|0\rangle $ and $|1\rangle $\ kets represent,
respectively, the vacuum and excited states. The first entry in each pair of
the kets is in Alice's possession and the second entry is in Rob's
possession. The subscirpts $\omega _{N}$ ($N=A,R$) of the kets reflect that
we consider the initial entanglement between these two modes and all the
rest modes are in vacuum states. Furthermore, we suppose that each player
has a device that is capabale of decting only the mode that she/he has
shared for the initial entanglement. Rob then moves with a constant
acceleration and Alice stays stationary. For accelerated observer the
suitable coordinates are Rindler coordinates. These coordinates define two
different regions, usually denoted by $I$ and $II$, and are called Rindler
regions. An observer in one region is causally disconnected from the other
region. In other words, an observer in region $I$ has no access to
information that leaks out to region $II$ and vice versa (for detail see
\cite{Alsing} and reference therein). A given Minskowski mode of a
particular frequency spreads over all positive Rindler frequencies that
peaks about the Minskowski frequency \cite{Takagi,Alsing2}. However, to
simplify our problem we consider a single mode only in the Rindler region $I$%
, which is valid if the observers' detectors are highly monochromatic that
detects the frequency $\omega _{A}\sim \omega _{B}=\omega $. From this point
onward, with this approximation, the frequency subscript of kets will be
dropped.

From the accelerated Rob's frame, the Minkowski vacuum state is found to be
a two-mode squeezed state \cite{Alsing},%
\begin{equation}
|0\rangle _{M}=\cos r|0\rangle _{I}|0\rangle _{II}+\sin r|1\rangle
_{I}|1\rangle _{II},  \label{2}
\end{equation}%
where $\cos r=\left( e^{-2\pi \omega c/a}+1\right) ^{-1/2}$. The constant $%
\omega $, $c$ and $a$, in the exponential stand, respectively, for Dirac
particle's frequency, the speed of light in vacuum and Rob's acceleration.
In Eq. (\ref{2}) the subscripts $I$ and $II$ of the kets represent the
Rindler modes in region $I$ and $II$, respectively. The excited state in
Minkowski spacetime is related to Rindler modes as follow \cite{Alsing}%
\begin{equation}
|1\rangle _{M}=|1\rangle _{I}|0\rangle _{II}.  \label{3}
\end{equation}

In terms of Minkowski modes for Alice and Rindler modes for Rob, the
maximally entangled initial state of Eq. (\ref{1}) by using Eqs. (\ref{2})
and (\ref{3}) becomes%
\begin{equation}
|\psi \rangle _{A,I,II}=\frac{1}{\sqrt{2}}\left( \cos r|0\rangle
_{A}|0\rangle _{I}|0\rangle _{II}+\sin r|0\rangle _{A}|1\rangle
_{I}|1\rangle _{II}+|1\rangle _{A}|1\rangle _{I}|0\rangle _{II}\right) .
\label{4}
\end{equation}%
Traditionally, we consider Rob to be in region $I$, then, he is causally
disconnected from region $II$. All the modes in region $II$ of Eq. (\ref{4})
needs to be discard. So, tracing over all the modes in region $II$ leaves
the following mixed density matrix between Alice and Rob%
\begin{eqnarray}
\rho _{A,I} &=&\frac{1}{2}[\cos ^{2}r|00\rangle _{A,I}\langle 00|+\cos
r(|00\rangle _{A,I}\langle 11|+|11\rangle _{A,I}\langle 00|)  \nonumber \\
&&\sin ^{2}r|01\rangle _{A,I}\langle 01|+|11\rangle _{A,I}\langle 11|].
\label{5}
\end{eqnarray}

\begin{table*}[htb]%
\caption{A single qubit Kraus operators for amplitude damping channel, phase
damping channel and depolarizing channel. \label{table:1}}%
\begin{tabular}{|c|c|}
\hline
amplitude damping & $E_{o}=\left(
\begin{array}{cc}
1 & 0 \\
0 & \sqrt{1-p_{1}}%
\end{array}%
\right) ,\qquad E_{1}=\left(
\begin{array}{cc}
0 & \sqrt{p_{1}} \\
0 & 0%
\end{array}%
\right) $ \\ \hline
phase damping & $E_{o}=\left(
\begin{array}{cc}
1 & 0 \\
0 & \sqrt{1-p_{2}}%
\end{array}%
\right) ,\qquad E_{1}=\left(
\begin{array}{cc}
0 & 0 \\
0 & \sqrt{p_{2}}%
\end{array}%
\right) $ \\ \hline
depolarizing & $%
\begin{array}{c}
E_{o}=\sqrt{1-p_{3}}\left(
\begin{array}{cc}
1 & 0 \\
0 & 1%
\end{array}%
\right) ,\qquad E_{1}=\sqrt{p_{3}/3}\left(
\begin{array}{cc}
0 & 1 \\
1 & 0%
\end{array}%
\right) , \\
E_{2}=\sqrt{p_{3}/3}\left(
\begin{array}{cc}
0 & -i \\
i & 0%
\end{array}%
\right) ,\qquad E_{3}=\sqrt{p_{3}/3}\left(
\begin{array}{cc}
1 & 0 \\
0 & -1%
\end{array}%
\right)%
\end{array}%
$ \\ \hline
\end{tabular}

\end{table*}%

\section{The system in noisy environment}

The effect of decoherence is studied via quantum channels in the Kraus
operators formalism. The Kraus operators for the three channels of a single
qubit system that we use in this paper are given in Table $1$. When a single
qubit system evolves under the action of amplitude damping channel there is
a probability to change the state of the system from state $|1\rangle $ to
state $|0\rangle $. When the density matrix of a single qubit system is
influenced by a phase damping channel, the diagonal elements remains
unaffected while the off-diagonal elements decay. On the other hand, the
study of the effect of a depolarizing noise on a quantum system is important
because a pure state of a system when influenced by such a noise is
replaced, with some probability, with a maximally mixed state. In Table $1$,
the action of the Kraus operators of each channel is parameterized by the
decoherence parameters $p_{i}$ ($i=1,2,3$) whose values lie between $0$ and $%
1$. For $p_{i}=0$, the channels have no effect on the system and for $%
p_{i}=1 $, the system is fully decohered. Furthermore, the Kraus operators
for each channel obey the completeness relation $\sum_{i}E_{i}^{\dag
}E_{i}=I $. The evolution of the initial density matrix of the system when
it is influenced by the global environment is given as follow%
\begin{equation}
\rho _{f}=\sum_{i}\sum_{j}\sum_{k}\left( E_{i}^{AR}E_{j}^{R}E_{k}^{A}\right)
\rho _{A,I}\left( E_{k}^{A\dag }E_{j}^{R\dag }E_{i}^{AR\dag }\right) ,
\label{6}
\end{equation}%
where $E_{k}^{A}=E_{m}^{A}\otimes I_{2}$, $E_{j}^{R}=I_{2}\otimes E_{n}^{R}$
are the Kraus operators of the local coupling of Alice's qubit and Rob's
qubit, respectively. The $I_{2}$ is identity operator and the subscripts $m$
and $n$ represent the number of single qubit Kraus operators of the
particular channel that is coupled to Alice's qubit or Rob's qubit. The $%
E_{i}^{AR}$ are the Kraus operators for the case of collective coupling and
are formed from all the possible combinations of the tensor product of the
Kraus operators of a single qubit channel in the form $E_{q}^{A}\otimes
E_{q}^{R}$. The subscripts $q$, represents the number of Kraus operators for
a one qubit channel used as a collective environment. The system is said to
be coupled with multilocal environment when both the qubits are influenced
by their local environments and it is said to be under the action of global
environment when both multilocal and collective couplings are switched on
simultaneously (global = multilocal + collective). First we consider the
situation in which Alice qubit is locally influenced by phase damping
channel, Rob's qubit is locally influenced by amplitude damping channel and
the two qubits are then coupled collectively to depolarizing channel. Under
such condition, the summations over $i$, $j$ and $k$ in Eq. (\ref{6}) are,
respectively, given by $i=0,1,2,...16$ and $j,k=0,1$. If $p_{1}$, $p_{2}$
and $p_{3}$ stand for decoherence parameters of amplitude damping channel,
phase damping channel and depolarizing channel, respectively, then the
non-zero elements of the final density matrix of the system in the light of
Eq. (\ref{6}) become%
\begin{eqnarray}
\rho _{11} &=&\frac{1}{36}%
(-8p_{3}^{2}(p_{1}-1)-(2p_{3}(4p_{3}-9)+9)(p_{1}-1)\cos 2r  \nonumber \\
&&-6p_{3}(p_{1}+1)+9(p_{1}+1)),  \nonumber \\
\rho _{14} &=&\frac{1}{18}(3-4p_{3})^{2}\sqrt{(-1+p_{1})(-1+p_{2})}\cos r,
\nonumber \\
\rho _{22} &=&\frac{1}{36}(9+8p_{3}^{2}(-1+p_{1})-9p_{1}+6p_{3}(1+p_{1})
\nonumber \\
&&+(9+2p_{3}(-9+4p_{3}))(-1+p_{1})\cos 2r),  \nonumber \\
\rho _{33} &=&\frac{1}{18}(p_{3}(9+4p_{3}(-1+p_{1})-15p_{1})+9p_{1}
\nonumber \\
&&+p_{3}(-3+4p_{3})(-1+p_{1})\cos 2r),  \nonumber \\
\rho _{41} &=&\frac{1}{18}(3-4p_{3})^{2}\sqrt{(-1+p_{1})(-1+p_{2})}\cos 2r,
\nonumber \\
\rho _{44} &=&\frac{1}{18}(9-9p_{1}+p_{3}(-9-4p_{3}(-1+p_{1})+15p_{1})
\nonumber \\
&&-p_{3}(-3+4p_{3})(-1+p_{1})\cos 2r).  \label{RA}
\end{eqnarray}%
The spin-flip matrix of the final density matrix of Eq. (\ref{6}) is defined
as $\tilde{\rho}_{f}=\left( \sigma _{2}\otimes \sigma _{2}\right) \rho
_{f}\left( \sigma _{2}\otimes \sigma _{2}\right) $, where $\sigma _{2}$ is
the Pauli matrix. The degree of entanglement in the two qubits mixed state
in a noisy environment can be quantified conveniently by concurrence $C$,
which is given as \cite{Wootter,Coffman}%
\begin{equation}
C=\max \left\{ 0,\sqrt{\lambda _{1}}-\sqrt{\lambda _{2}}-\sqrt{\lambda _{3}}-%
\sqrt{\lambda _{4}}\right\} \qquad \lambda _{i}\geq \lambda _{i+1}\geq 0,
\label{7}
\end{equation}%
where $\lambda _{i}$ are the eigenvalues of the matrix $\rho _{f}\tilde{\rho}%
_{f}$. The eigenvalues under the action of global environment become%
\begin{eqnarray}
\lambda _{1,2} &=&\frac{1}{324}%
(54p_{3}-36p_{3}^{2}+81p_{1}-216p_{3}p_{1}+144p_{3}^{2}p_{1}-81p_{1}^{2}+216p_{3}p_{1}^{2}
\nonumber \\
&&-144p_{3}^{2}p_{1}^{2}+(3-4p_{3})^{2}(-1+p_{1})(-24p_{3}(-1+p_{2})+16p_{3}^{2}(-1+p_{2})
\nonumber \\
&&+9(-2+p_{1}+p_{2}))\cos
^{2}r+2(3-4p_{3})^{2}p_{3}(-3+2p_{3})(-1+p_{1})^{2}\cos ^{4}r  \nonumber \\
&&\pm 2[(3-4p_{3})^{4}(-1+p_{1})(-1+p_{2})\cos ^{2}r(-9(-6p_{3}(1-2p_{1})^{2}
\nonumber \\
&&+4p^{2}(1-2p_{1})^{2}+9(-1+p_{1})p_{1})+9(3-4p_{3})^{2}(-1+p_{1})^{2}\cos
^{2}r  \nonumber \\
&&+2(3-4p_{3})^{2}p_{3}(-3+2p_{3})(-1+p_{1})^{2}\cos ^{4}r))]^{1/2},
\nonumber \\
\lambda _{3} &=&\lambda _{4}=\frac{1}{1296}%
(3(-96p_{3}^{3}(-1+p_{1})^{2}+32p_{3}^{4}(-1+p_{1})^{2}-54(-1+p_{1})p_{1}
\nonumber \\
&&-6p_{3}^{2}(-7+14p_{1}+p_{1}^{2})+9p_{3}(5-10p_{1}+13p_{1}^{2}))+2(3-4p_{3})^{2}
\nonumber \\
&&(-1+p_{1})(-6p_{3}(-1+p_{1})+4p_{3}^{2}(-1+p_{1})+9p_{1})\cos 2r  \nonumber
\\
&&+(3-4p_{3})^{2}p_{3}(-3+2p_{3})(-1+p_{1})^{2}\cos 4r),  \label{8}
\end{eqnarray}%
where in $\lambda _{1,2},$ the two eigenvalues are differentiated by $\pm $
sign, respectively. Using Eq. (\ref{7}) the concurrence can
straightforwardly be calculated. It is important to note that the
concurrence defined by the eigenvalues of Eq. (\ref{8}) reduces to the
result of Ref. \cite{Alsing} when the decoherence parameters $p_{i}=0$.
\begin{figure}[h]
\begin{center}
\begin{tabular}{cc}
\includegraphics[scale=0.6]{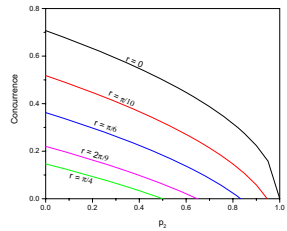} \put( -480,150) \ \ (a) &
\end{tabular}
\ \ \ \ \ \ \
\begin{tabular}{cc}
\includegraphics[scale=0.6]{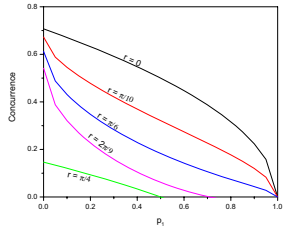} \put(-480,150) \ \ \ (b) &
\end{tabular}%
\end{center}
\caption{(color online) (a) The concurrence for different values of the
acceleration is plotted against the decoherence parameter $p_{2}$, for the
case when both the qubits are coupled to their respective environments in
multilocal coupling. The other parameters are set to $p_{1}=0.5$, and $%
p_{3}=0$. (b) The concurrence against $p_{1}$ for $p_{2}=0.5$ and $p_{3}=0$
for different values of the acceleration of Rob's frame.}
\label{Figure1ab}
\end{figure}

To see how the concurrence and hence the entanglement change when the system
is coupled to noisy environments in the presence of Unruh effect, we plot
the concurrence for each kind of the coupling of the three channels against
the decoherence parameters for various values of the acceleration $r$ of the
noninertial frame. In Fig. $1$($a,b$), the concurrence, when the collective
environment is switched off ($p_{3}=0$), is plotted for the case of
multilocal coupling. Fig. $1a$ shows the behavior of concurrence against the
decoherence parameter $p_{2}$ (phase damping) for $p_{1}=0.5$ (amplitude
damping). One can see that the concurrence is strongly acceleration
dependent as the initial value of the concurrence decreases with increased
acceleration. For a given acceleration, the decoherence parameter
considerably damps the concurrence as the system evolves and it becomes zero
at a definite value of the decoherence parameter (particular time) and hence
loss of the entanglement occurs. The ESD occurs earlier for the case of
larger acceleration. On the other hand, the behavior of concurrence is
different when it is plotted against $p_{1}$ for a particular value of $%
p_{2} $. Fig. $1b$ shows such a situation for $p_{2}=0.5$. Here the
concurrence is not as much strongly acceleration dependent as in the case of
Fig. $1a$, except for the larger values of acceleration. For smaller
acceleration of the noninertial frame, the ESD does not happen, the
concurrence goes to zero only when the channel is fully decohered (infinite
time). However, for large values of the acceleration ($r=\pi /4$), the
concurrence is influenced identically by both $p_{1}$ and $p_{2}$. From the
different behavior of concurrence and hence of ESD in the range of smaller
acceleration, the two channels can be used to identify the stationary and
the accelerated observers. The behavior of concurrence is shown in Fig. $2$ (%
$a,b$) for the case when both the collective and multilocal environments are
simultaneously switched on, that is, when the system is coupled to global
environment. In Fig. $2a$, the values of decoherence parameters that
parameterize the multilocal environment are set to $p_{1}=p_{2}=0.1$. It can
be seen that the concurrence is strongly dependent both on acceleration and
decoherence parameter. The ESD occurs faster than in the case when the
system is coupled to multilocal environment only. Fig. $2b$ shows the
situation in which $p_{1}=p_{2}=p_{3}=p$. One can see from the figure that
under such constraint, the concurrence, as compared to Fig. $2a$, becomes
more dependent on decoherence parameter and less dependent on the
acceleration.
\begin{figure}[h]
\begin{center}
\begin{tabular}{cc}
\includegraphics[scale=0.6]{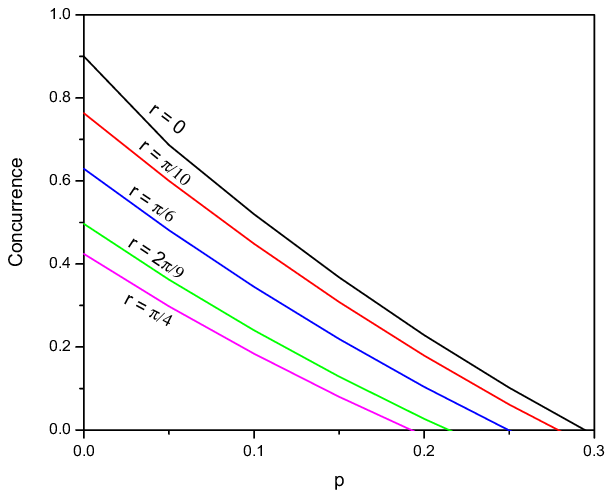} \put( -480,150) \ \ (a) &
\end{tabular}
\ \ \ \ \ \ \
\begin{tabular}{cc}
\includegraphics[scale=0.6]{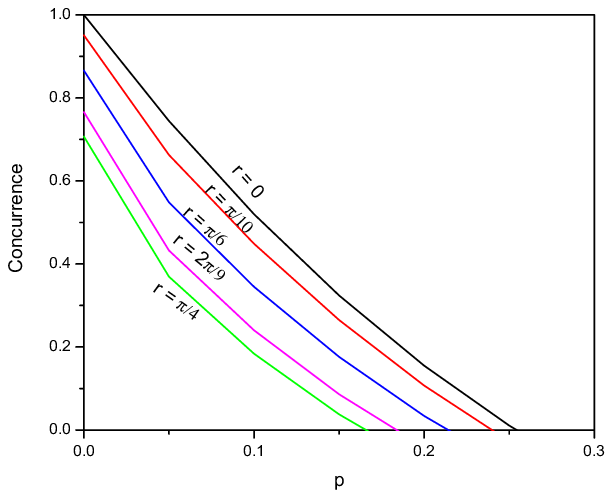} \put(-480,150) \ \ \ (b) &
\end{tabular}%
\end{center}
\caption{(color online) (a) The concurrence for different values of the
acceleration is plotted against the decoherence parameter $p_{3}$, for the
case when both the multilocal and collective environments are switched on.
The other parameters are set to $p_{1}=p_{2}=0.1$. (b) The concurrence is
plotted for different values of the acceleration against decoherence
parameter $p$ under the conditions $p_{1}=p_{2}=p_{3}=p$.}
\label{Figure2ab}
\end{figure}

Now, we want to see the effect of interchanging multilocal channels that is,
when Alice's qubit is locally coupled to an amplitude damping environment
and Rob's qubit interacts locally with a phase damping environment. For this
case, the summations over $i,j$ and $k$ in Eq. (\ref{6}) remain unchanged.
The concurrence can straightforwardly be calculated as in the previous case.
Without writing its mathematical form, we prefer to show its behavior by
plotting it against the decoherence parameters. The concurrence as a
function of $p_{1}$ for $p_{2}=0.5$ is shown in Fig. $3$ for different
values of acceleration when the collective environment is switched off. The
figure shows that the concurrence is significantly both acceleration and
decoherence parameter dependent such that it decreases with the increasing
values of both of these quantities. However, no ESD occurs for any value of
the acceleration, the concurrence goes to zero for every acceleration only
when $p_{1}=1$. Unlike the previous case, the plot of concurrence against $%
p_{2}$ is exactly the same as given in Fig. $3$. This means that the rapid
fall of concurrence in the previous case (Fig. $1a$) is somehow compensated
by the acceleration of the moving frame in this case. The two channels are
not differentiable in this case. Nevertheless, ESD does occur in this case
when the system evolves in the global environment. The effect of global
environment on the concurrence is shown in Fig. $4$. A comparison with Fig. $%
2a$ shows that the concurrence is influenced almost identically in the two
cases by the decoherence parameters, however, it is less affected by the
acceleration in the later case.
\begin{figure}[h]
\begin{center}
\begin{tabular}{ccc}
\vspace{-0.5cm} \includegraphics[scale=1.2]{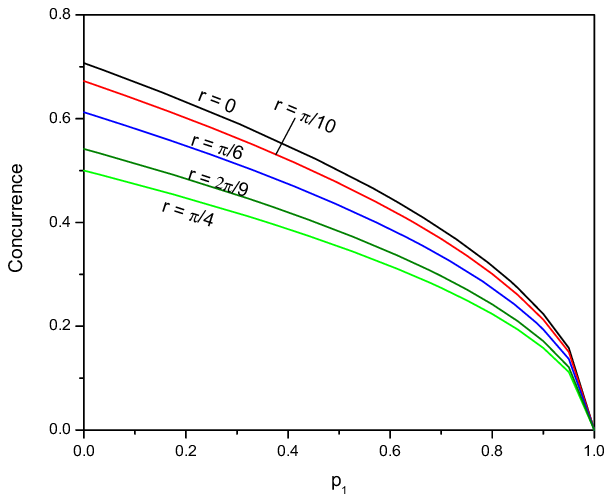}\put(-100,200) &  &
\end{tabular}%
\end{center}
\caption{(color online) The concurrence for different values of the
acceleration is plotted against the decoherence parameter $p_{1}$, for the
case of multilocal coupling when Alice's qubit is coupled to amplitude
damping channel and Rob's qubit to phase damping channel. The other
parameters are set to $p_{2}=0.5$, and $p_{3}=0$.}
\label{Figure3}
\end{figure}

\begin{figure}[h]
\begin{center}
\begin{tabular}{ccc}
\vspace{-0.5cm} \includegraphics[scale=1.2]{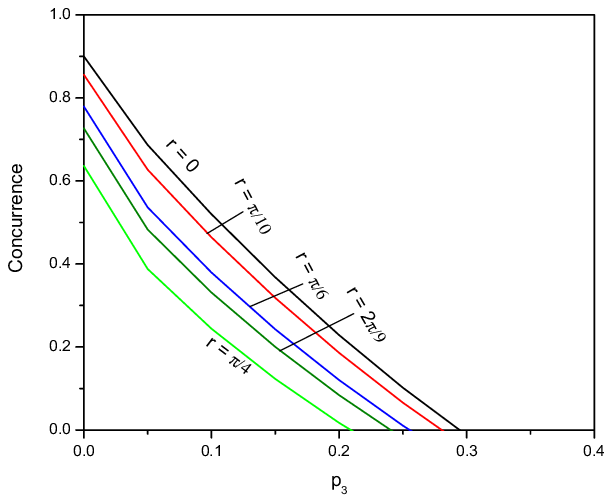}\put(-100,170) &  &
\end{tabular}%
\end{center}
\caption{(color online) The concurrence for different values of the
acceleration is plotted against the decoherence parameter $p_{3}$, for the
case of global environment. The other parameters are set to $p_{1}=p_{2}=0.1$%
.}
\label{Figure4}
\end{figure}

Next, we consider the setup in which Alice's qubit interacts locally with a
depolarizing channel and Rob's qubit interacts locally with a phase damping
channel as before. The two qubits then interact collectively with an
amplitude damping channel. The summations over $i$, $j$ and $k$ in Eq. (\ref%
{6}) become $i=k=0,1,2,3$ and $j=0,1$. Using Eq. (\ref{6}) the final density
matrix can easily be calculated. After finding the spin-flip matrix, the
four eigenvalues of the matrix $\rho \tilde{\rho}$ are given by
\begin{figure}[h]
\begin{center}
\begin{tabular}{cc}
\includegraphics[scale=0.6]{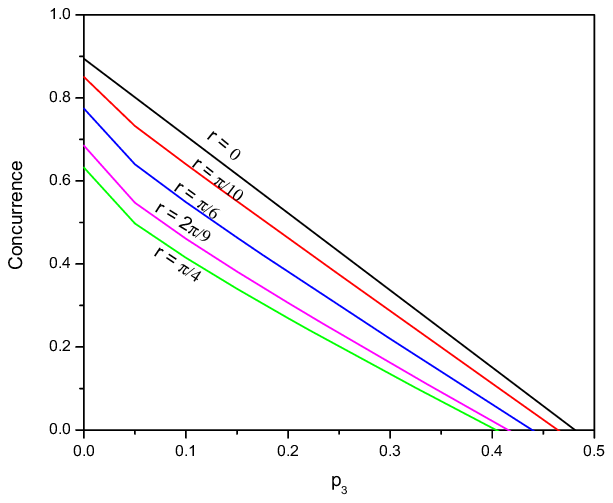} \put( -480,150) \ \ (a) &
\end{tabular}
\ \ \ \ \ \ \
\begin{tabular}{cc}
\includegraphics[scale=0.6]{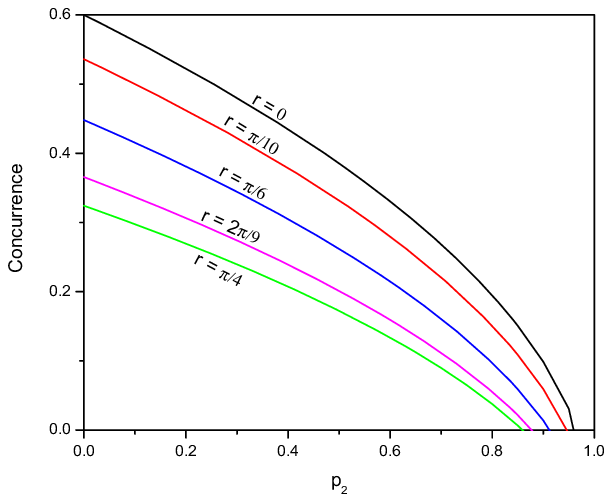} \put(-480,150) \ \ \ (b) &
\end{tabular}%
\end{center}
\caption{(color online) (a) The concurrence for different values of the
acceleration is plotted against the decoherence parameter $p_{3}$, for the
case when both the qubits are coupled to their respective environments in
multilocal coupling. The other parameters are set to $p_{2}=0.2$, and $%
p_{1}=0$. (b) The concurrence against $p_{2}$ for $p_{3}=0.2$ and $p_{1}=0$
for different values of the acceleration of Rob's frame.}
\end{figure}
\begin{eqnarray}
\lambda _{1,2} &=&\frac{1}{36}%
(-1+p_{1})^{2}(18+9p_{1}^{2}-9p_{2}-36p_{3}+12p_{1}p_{3}  \nonumber \\
&&-12p_{1}^{2}p_{3}+24p_{2}p_{3}+20p_{3}^{2}-8p_{1}p_{3}^{2}+4p_{1}^{2}p_{3}^{2}-16p_{2}p_{3}^{2}
\nonumber \\
&&\pm \sqrt{2}[-(-1+p_{2})(3-4p_{3})^{2}\cos ^{2}r(-3+p_{3}+p_{3}\cos 2r)
\nonumber \\
&&\times (-3(1+p_{1}+2p_{1}^{2})+2(-1+p_{1})^{2}p_{3}+(-1+p_{1})  \nonumber
\\
&&\times (3+2(-1+p_{1})p_{3})\cos 2r)]^{1/2}+(9(-2+p_{1}+p_{2})  \nonumber \\
&&+6(7-3p_{1}+2p_{1}^{2}-4p_{2})p_{3}-8(3-2p_{1}+p_{1}^{2}-2p_{2})p_{3}^{2})%
\sin ^{2}r  \nonumber \\
&&+2(-1+p_{1})p_{3}(3+2(-1+p_{1})p_{3})\sin ^{4}r),  \nonumber \\
\lambda _{3} &=&\lambda _{4}=\frac{1}{144}%
(-1+p_{1})^{2}(3(2p_{3}^{2}(-1+p_{1})^{2}+6p_{1}(1+2p_{1})  \nonumber \\
&&+p(1+7p_{1}-8p_{1}^{2}))+2(4p_{3}^{2}(-1+p_{1})^{2}-9p_{1}  \nonumber \\
&&-12p(-1+p_{1})p_{1})\cos 2r+p_{3}(3+2p_{3}(-1+p_{1}))(-1+p_{1})\cos 4r).
\label{Eig2}
\end{eqnarray}%
\begin{figure}[h]
\begin{center}
\begin{tabular}{ccc}
\vspace{-0.5cm} \includegraphics[scale=1.2]{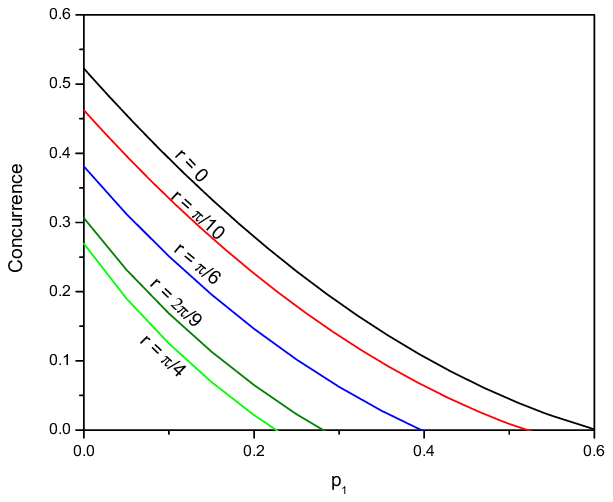}\put(-100,170) &  &
\end{tabular}%
\end{center}
\caption{(color online) The concurrence for different values of the
acceleration is plotted against the decoherence parameter $p_{1}$, for the
case of global environment. The other parameters are set to $p_{2}=p_{3}=0.2$%
.}
\end{figure}
The concurrence can be found by using these eigenvalues in Eq. (\ref{7}). To
study the influence of the noisy environment on the concurrence, we proceed
by plotting it against the decoherence parameters as done in the previous
cases. First we consider the effects of decoherence in the multilocal
coupling. The plot of concurrence against $p_{3}$ ($p_{2}=0.2$) and $p_{2}$ (%
$p_{3}=0.2$) for $p_{1}=0$ is shown in Fig. $5$($a,b$). Though the behavior
of concurrence is different in both of these figures, however, the ESD
cannot be avoided. The ESD in this case occurs faster as compared to the
previous two cases. On the other hand, the damping of concurrence due to
acceleration is stronger in Fig. $5b$ than in Fig. $5a$. Finally we show the
behavior of concurrence when the system is coupled with the global
environment. Fig. $6$ shows the concurrence of the system against $p_{1}$
for different values of acceleration with $p_{2}=p_{3}=0.2$. It is obvious
from the last two figures that when one qubit is locally coupled to
depolarizing channel there is no way to avoid ESD and it happens faster than
the other two cases.

\section{Conclusion}

In conclusion, we have studied the behavior of entanglemnent of Dirac fields
by considering different combinations of the three different quantum
channels. We show that under multilocal environment when Alice's qubit is
coupled to a phase damping channel and Rob's qubit is coupled to an
amplitude damping channel, the ESD may or may not occur. However, the
distinguishing effects of the two channels may be used to identify which of
the observer is in motion. On the other hand, if the multilocal coupling of
the two channels with two qubits is interchanged, the two channels affect
the concurrence identically and cannot be distinguished from each other.
However, the ESD can be avoided. The concurrence goes to zero only when one
of the channels is fully decohered. Furthermore, it is shown that
irrespective of the multilocal coupling's configuration, the ESD occurs when
the collective environment is present. Moreover, It is shown that ESD occurs
faster and can not be avoided when one of the two qubits is locally coupled
to the depolarizing channel.

\end{document}